\documentclass[twocolumn,showpacs,preprintnumbers,amsmath,amssymb]{revtex4}

\usepackage{graphicx}
\usepackage{dcolumn}
\usepackage{bm}

\begin{document}

\title{Exact dynamics of quantum correlations of two qubits coupled to bosonic baths}

\author{Chen Wang$^{1}$}
\author{Qing-Hu Chen$^{1,2,}$}\email{qhchen@zju.edu.cn}

\address{
$^{1}$ Department of Physics, Zhejiang University, Hangzhou 310027,
P. R. China \\
$^{2}$ Center for Statistical and Theoretical Condensed Matter
Physics, Zhejiang Normal University, Jinhua 321004, P. R. China
 }
\date{\today}

\begin{abstract}
The counter rotating-wave term (CRT) effects from the system-bath
coherence on the dynamics of quantum correlation of two qubits in
two independent baths and a common bath are systematically
investigated. The hierarchy approach is extended to  solve  the
relevant spin boson models with the Lorentzian spectrum, the exact
dynamics for the  quantum entanglement and quantum discord (QD) are
evaluated, and the comparisons with previous ones under the
rotating-wave approximation  are performed. For the two independent
baths, beyond the weak system-bath coupling, the CRT essentially
changes the evolution of  both entanglement and QD. With the
increase of the coupling, the revival of the entanglement is
suppressed dramatically and finally disappears,  and the QD becomes
smaller monotonically. For the common bath, the entanglement is also
suppressed by the CRT, but the QD  shows quite different behaviors,
if initiated from the correlated Bell  states. In the non-Markovian
regime, the QD is almost not influenced by the CRT and generally
finite in the long time evolution at any coupling, while in the
Markovian regime, it is significantly enhanced with the strong coupling.

\end{abstract}


\maketitle

\section{introduction}

Quantum correlation, originated from the coherent superposition in
quantum mechanics~\cite{einstein1,bell1}, has been attracting
persistent attention ranging from quantum information, quantum
optics, to many-body physics~\cite{nielsen1,taylor1,modi1}. As the
characteristic trait of the quantum correlation proposed by
Schr\"{o}dinger~\cite{schrodinger1} without the classical
counterpart, quantum entanglement played the crucial role in the
highly efficient quantum computation and quantum information
processing~\cite{vidal1,horodechi1}. It has also been applied to
identify the phase transition in quantum many-particle
systems~\cite{amico1}. However, there exist exceptions where the
entanglement seems unnecessary, such as in the Grover search
algorithm~\cite{grover1,ahn1} and the deterministic quantum
computation with one pure qubit (DQC$1$)\cite{knill1}. It is evident
that other kind of the nonclassical quantum correlation rather
entanglement dominates the computational implementation of
DQC$1$~\cite{lanyon1}. Zurek and Vedral described one particular measure to
quantify all nonclassical correlation, termed as quantum discord
(QD)~\cite{ollivier1,zurek1,henderson1}. Only when the QD
disappears, the information can be safely obtained by locally
measuring the distantly separate subsystem without disturbing the
bipartite quantum system, where the corresponding state is fully
classical. Recently, the QD has successfully been used to study the
quantum phase transition of the critical
systems~\cite{werlang,dillenschneider1,maziero1,allegra1,chen,wang},
operational meanings for quantum processors~\cite{gu1}, and state
preparation~\cite{walther1}.

It is known that the realistic system inevitably interacts with the
environment, resulting in the decoherence, which is of fundamental
importance in the quantum information processing and
measurement~\cite{breuer1,weiss1}. Specially, Yu and Eberly
~\cite{yu1} found that the dynamical behavior of the entanglement is
significantly different from the single qubit in a pair of qubits
separately coupling with Markovian baths, which can be modeled from the
well-known spin boson models~\cite{leggett1}. It shows finite time
disentanglement called as entanglement sudden death
(ESD)~\cite{yu1}. However, this was investigated under quantum
master equation with the Born and Markov approximation, which
assumes that system-bath interaction is weak and the relaxation of
the bath is much faster than that of the system. Then further works
have been done in the non-Markovian regime in the framework of the
rotating-wave approximation (RWA) to reveal novel characters of the
quantum correlations compared to the Markovian
effect~\cite{bellomo1,bellomo2,maniscalco1,werlang1,wang1,franco1},
which is only valid under the condition of weak system-bath interaction.
Besides, entanglement dynamics beyond the weak coupling spin boson
models has also been exploited based on the optimal polaron
transformation~\cite{cao1}, where the higher order terms are
neglected. It is pointed out that  the corresponding accuracy  in
the wide parameter zone should be carefully analyzed~\cite{lee1}. Therefore,
It is highly desirable to solve the dynamics without performing any approximation.

Beyond both the Born-Markov approximation and RWA, Tanimura \emph{et al.}~\cite{tanimura1}
have developed an efficient hierarchy approach,
which has been later extensively applied to some chemical and
biophysical systems~\cite{ishizaki1,xu1}. Recently, this method has
also been  extended to study  the  dynamics of entanglement for two
qubits  coupled to a common  bath~\cite{arend1,ma1}.  However, the
dynamics of QD,  one recently mostly studied quantum correlations
has not been investigated  beyond the Born-Markov
approximation and the RWA. Moreover,  the exact study of the dynamics of
both entanglement and QD for  two  qubits  coupled to
their own baths  are not available in the literature, to the best of
our knowledge. This case is designed for two remote qubits each
interacting locally with its own environment. Such a physical
condition is also relevant to the quantum information science and
quantum computation.

In the present paper, to give a comprehensive picture of the two
qubits coupled to their own bath and common bath, we will extend the
hierarchy equation to these two kinds of baths to study the dynamics
of the pairwise entanglement and the QD in both Markovian
and non-Markovian regimes. The effect of the counter-rotating term
on these dynamics in a wide coupling regime will be systematically
studied. This paper is organized as follows. In Sec. II we briefly
review the definition of the  QD. In Sec. III we exactly
solve the reduced qubits density matrix of two independent spin
boson models, and investigate its quantum correlation compared with
that in RWA. In Sec. IV the quantum correlation of two separate
qubits interacting with a common bath is analyzed. Finally, a summary is given in Sec. V.

\section{ Quantum Discord }

The  QD  is one main route to fully measure the
nonclassical correlation, which can be extended from the classical
information theory~\cite{ollivier1,zurek1,henderson1,modi1}. It is
interpreted as the difference of two quantum mutual correlations of
subsystems $\mathcal{A}$ and $\mathcal{B}$, before and after local
measurement operated on subsystem
$\mathcal{B}$~\cite{ollivier1,zurek1}. The total quantum correlation
of two subsystems is determined by their joint density matrix $\rho$
as
$\mathcal{I}(\rho)=S(\rho_{\mathcal{A}})+S(\rho_{\mathcal{B}})-S(\rho)$,
where
$\rho_{\mathcal{A}(\mathcal{B})}=\textrm{Tr}_{\mathcal{B}(\mathcal{A})}\{\rho\}$.
The von Neumann entropy $S(\rho_a)=-\textrm{Tr}\{\rho_a\ln\rho_a\}$.
The other quantum mutual information $\mathcal{J}(\rho)$, which is
equivalent to $\mathcal{I}(\rho)$ in classical information frame, is
derived by locally measuring  subsystem $\mathcal{B}$ with a
complete set of orthonormal projectors $\{\Pi^{\mathcal{B}}_k\}$,
where $k$ is one outcome state of $\mathcal{B}$. After this
measurement, the joint density is reduced to conditional counterpart
as
$\rho_k=[(\mathbb{I}{\otimes}\Pi^{\mathcal{B}}_k)\rho(\mathbb{I}{\otimes}\Pi^{\mathcal{B}}_k)]/p_k$,
where the corresponding probability is
$p_k=\textrm{Tr}\{(\mathbb{I}{\otimes}\Pi^{\mathcal{B}}_k)\rho\}$.
Then the mutual information based on specific performance is shown
as $\mathcal{Q}=S(\rho_{\mathcal{A}})-\sum_kp_kS(\rho_k)$. Hence, we
should maximize it to capture all classical correlation by obtaining
$\mathcal{J}(\rho)=\max_{\{\Pi^{\mathcal{B}}_k\}}\{\mathcal{Q}\}$.
Finally, the  QD  is defined as
\begin{eqnarray}
\mathcal{D}(\rho)=\mathcal{I}(\rho)-\mathcal{J}(\rho).
\end{eqnarray}
The quantum nature of the system can be explicitly observed from the
 QD. It is only zero in pure classical correlation
condition, and has the same values as the entanglement in pure state.
While it remains finite even in separable mixed states, where the
corresponding entanglement may completely
disappear~\cite{ollivier1}.

\section{two independent spin-boson models}
The Hamiltonian of two independent qubits coupled two independent
bosonic baths $A$ and $B$ is given by   $H=H_s+H_b+H_{sb}$, reads
\begin{eqnarray}~\label{tisp1}
H_s&=&\sum_{a=A,B}\frac{\omega_a}{2}\sigma^a_z, \nonumber\\
H_b&=&\sum_{a=A,B;k}\omega^a_kb^{\dag}_{a,k}b_{a,k}, \nonumber\\
H_{sb}&=&\sum_{a=A,B;k}\sigma^a_x(g^a_kb_{a,k}+g^{a*}_kb^{\dag}_{a,k}).
\end{eqnarray}
where $a=A, B$, $\sigma^a_{\beta}~(\beta=x,y,z)$ is the Pauli operator with Zeeman
energy $\omega_a$ of spin $\beta$, $b^{\dag}_{a,k}~(b_{a,k})$
creates (annihilates) one boson with frequency $\omega^a_k$ in bath
$a$, and  $g^{a}_k$ is the coupling strength between the system and the
bath.

The spectra density of the bosonic bath is selected as the following Lorentz distribution
\begin{eqnarray}~\label{ld}
J_a(\omega)=\frac{1}{2\pi}\frac{\lambda_a\gamma^2_a}{(\omega-\omega^a_0)^2+\gamma_a^2}.
\end{eqnarray}
which can describe an bosonic field inside an imperfect cavity mode $\omega_0^a$ with
the system-bath coupling strength $\lambda_a$.
The corresponding correlation function at zero temperature is shown as~(\ref{app15})
\begin{eqnarray}~\label{tisp2}
C_{aa}(\omega)=\frac{\lambda_a\gamma_a}{2}\exp[-(\gamma_a+i\omega^a_0)t].
\end{eqnarray}
Then by applying the Feynman-Vernon influence functional procedure
in Appendix, the hierarchy equation is derived by~(\ref{app19})
\begin{eqnarray}~\label{tisp3}
&&\frac{\partial}{{\partial}t}\rho_{\vec{n},\vec{m}}(t) \\
&=&-(iH^{\times}_S+\vec{n}\cdot\vec{\mu}_A+\vec{m}\cdot\vec{\mu}_B)\rho_{\vec{n},\vec{m}}(t) \nonumber\\
&&+\sum_{k=1,2}(-1)^kQ^{\times}_A\rho_{\vec{n}+\vec{e}_k,\vec{m}}(t) \nonumber\\
&&+\frac{\lambda_A\gamma_A}{4}\sum_{k=1,2}n_k[Q^{o}_A+(-1)^{k+1}Q^{\times}_A]\rho_{\vec{n}-\vec{e}_k,\vec{m}}(t) \nonumber\\
&&+\sum_{k=1,2}(-1)^kQ^{\times}_B\rho_{\vec{n},\vec{m}+\vec{e}_k} \nonumber\\
&&+\frac{\lambda_B\gamma_B}{4}\sum_{k=1,2}m_k[Q^{o}_B+(-1)^{k+1}Q^{\times}_B]\rho_{\vec{n},\vec{m}-\vec{e}_k}(t).\nonumber
\end{eqnarray}
The commutation relations $A^{\times}B=AB-BA$ and $A^{o}B=AB+BA$.
The unit vectors $\vec{e}_1=(1,0)$, $\vec{e}_2=(0,1)$, and $\vec{n}=(n_1,n_2)$,
$\vec{m}=(m_1,m_2)$. $Q_a~(a=A,B)$ in the system-bath interaction represents $\sigma^a_x$.
At the initial state, only $\rho_{(0,0),(0,0)}(0)$ is not zero. For the simplest condition in the following calculations,
the two subsystems are identical as $\omega_a=\omega_0$, $\lambda_a=\lambda$, $\gamma_a=\gamma$.

For comparison, we thereby briefly review the main results derived in the RWA,
where system-bath interaction becomes $H^R_{sb}=\sum_{a,k}(g^a_k{\sigma}^a_+b_{a,k}+g^{a*}_k{\sigma}^a_-b^{\dag}_{a,k})$.
When the initial system reduced density matrix in the  basis $\{|1{\rangle}=|11{\rangle},|2{\rangle}=|10{\rangle},
|3{\rangle}=|01{\rangle},|4{\rangle}=|00{\rangle}\}$ is $X$ form
\begin{eqnarray}~\label{tisp4}
\rho_{AB}(0)=
\begin{pmatrix}
\rho_{11}(0) & 0 & 0 & \rho_{14}(0) \\
0 & \rho_{22}(0) & \rho_{23}(0) & 0 \\
0 & \rho_{32}(0) & \rho_{33}(0) & 0 \\
\rho_{41}(0) & 0 & 0 & \rho_{44}(0)
\end{pmatrix},
\end{eqnarray}
then we have
\begin{eqnarray}~\label{tisp5}
\rho_{11}(t)&=&\rho_{11}(0)P^2_t,~\rho_{22}(t)=\rho_{22}(0)P_t+\rho_{11}(0)P_t(1-P_t), \nonumber\\
\rho_{33}(t)&=&\rho_{33}(0)P_t+\rho_{11}(0)P_t(1-P_t),\nonumber\\
\rho_{44}(t)&=&1-[\rho_{11}(t)-\rho_{22}(t)-\rho_{33}(t)], \nonumber\\
\rho_{14}(t)&=&\rho_{14}(0)P(t),~\rho_{23}(t)=\rho_{23}(0)P(t),
\end{eqnarray}
where
\begin{equation}
P_t=e^{-{\gamma}t}[\cos(Rt)+\frac{\gamma}{2R}\sin(Rt)]^2
\end{equation}
with $R=\sqrt{\lambda\gamma/2-\gamma^2/4}$. It is known that the
bath relaxation time is $\tau_b{\approx}1/\gamma$, and the
system-bath correlation time is $\tau_r{\approx}1/\lambda$. From the time dependence of
the function $P_t$, we immediately know that the Markovian regime corresponds to
$\gamma/2>\lambda$ where $P_t$ decays exponentially, while
the non-Markovian regime to $\gamma/2<\lambda$ where  $P_t$ shows
oscillating behavior describing the coherent process between the system and the bath.
This non-Markovian condition has been realized
in cavity quantum electrodynamics having Rydberg atoms inside the
Fabry-Perot cavities with $\gamma/\lambda{\approx}0.1$~\cite{kuhr1}.
We should point out here that the Markovian and non-Markovian regimes are only
defined in the framework of the RWA, which can not be well defined
in the exact treatment without the RWA. Throughout the present paper, we still use such notations.
Doing so is only for comparisons and analysis.

As usual, the  initial Bell states with anti-correlated spins and
correlated spins will be considered in the present paper
\begin{eqnarray}
|\Phi(\alpha){\rangle}&=&\alpha|10{\rangle}+\sqrt{1-\alpha^2}|01{\rangle}, ~\label{tisp61}\\
|\Psi(\alpha){\rangle}&=&\alpha|00{\rangle}+\sqrt{1-\alpha^2}|11{\rangle},~\label{tisp62}
\end{eqnarray}
which both obey $X$ form in Eq.~(\ref{tisp4}) during the evolution. So
it is straightforward to derive the concurrence, a pairwise
entanglement\cite{bellomo1}
\begin{eqnarray}~\label{tisp7}
C_{\Phi}(t)&=&\max\{0,2\alpha\sqrt{1-\alpha^2}P_t\}, \\
C_{\Psi}(t)&=&\max\{0,2\sqrt{1-\alpha^2}P_t[\alpha-\sqrt{1-\alpha^2}(1-P_t)]\}.\nonumber\\
\end{eqnarray}
The analytical expression for  QD  in this case is quite
complicated and also only consists of $\alpha$ and $P_t$, which is
implied in Refs. \cite{wang1,fanchini1,wang0}.

It is shown analytically above that the dynamics of both
entanglement and  QD  under the RWA can be distinguished in the
non-Markovian and  Markovian regimes, characterized by the ratio
$\gamma/\lambda$. In each regime, the essential feature only depends
on the weights $\alpha^2$ of the initial states, and independent of
the system-bath coupling strength $\lambda$.  In other words, in the
framework of the RWA, all essential properties for the physical
observable are independent of the system-bath coupling strength,
which is not always true with the increasing coupling realized in
many recent experiments~\cite{niemczyk1}. Hence, it is necessary to go
beyond the RWA.

\subsection{Exact dynamics and comparisons}

\begin{figure}[tbp]
\includegraphics[scale=0.4]{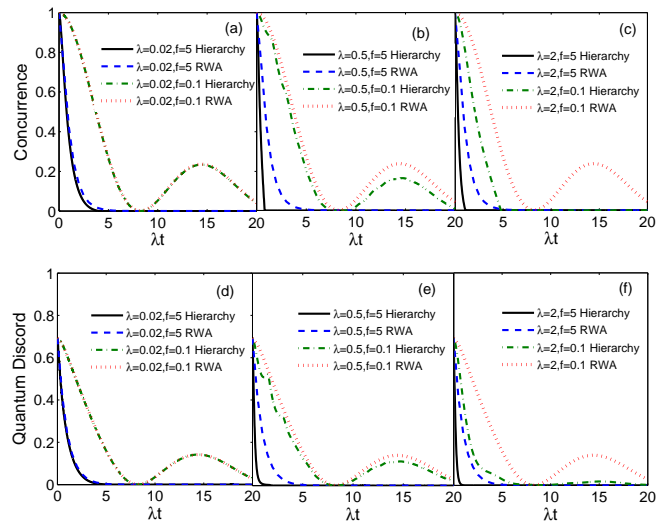}
\caption{Dynamics of quantum correlation of two qubits coupled to
two independent baths with initial anti-correlated Bell state
$|\Phi(\frac{1}{\sqrt{2}}){\rangle}$. $\omega_0=1$,
$\gamma=f\lambda$. } \label{TATBcompare}
\end{figure}

The dynamics of the pairwise concurrence of the two qubits in their
own baths without the RWA  can be obtained exactly by the hierarchy
approach outlined above.  Initiated from the anti-correlated  Bell
state $(\ref{tisp61})$ with $\alpha=1/\sqrt{2}$, the time evolution
of the concurrence from weak, intermediate, to strong coupling cases
are presented in the upper panel of Fig.~\ref{TATBcompare}. The
concurrence under the RWA has been investigated exactly  in
Ref.~\cite{bellomo1} both in the Markovian and non-Markovian
regimes. The relevant results for the same parameters are also
reproduced in the upper panel of Fig.~\ref{TATBcompare} for
comparison.

In the weak system-bath coupling ($\lambda=0.02$), the rotating-wave
term dominates the entanglement evolution as indicated in
Fig.~\ref{TATBcompare}(a), and the results including the CRT show
negligible deviation from those in RWA. In the non-Markovian regime,
the concurrence exhibits periodic oscillations with amplitude
damping. The corresponding time of the zero points is
$t_n=[n\pi-\arctan(R/2\gamma)]/R~(n=1,2,\cdots)$. Whether it only
vanishes at these discrete  moments is very subtle, the periods of
the time for zero entanglement is too short to be visible in this
case  . While in the Markovian regime, the concurrence decays
exponentially and vanish asymptotically. The decay rate is larger
than the  non-Markovian one. So at such a weak coupling, we have not
found any evident effect from the CRT, so the previous RWA
description is really valid.

What happens if we increase the system-bath coupling to the
intermediate regime, say $\lambda=0.5$. It is interesting to observe
in Fig.~\ref{TATBcompare}(b) for $\lambda=0.5$  that  the dynamic
behavior is evidently different from that in the RWA. From Eq. (11),
one can see that if initiated from anti-correlated Bell state, in
the RWA the ESD never happens, which was  first found in
Ref.~\cite{bellomo1} in the same system. But without the RWA, in both
the Markovian ($f=5$) and the non-Markovian ($f=0.1$) regimes, the
concurrence  exhibits ESD obviously. It follows that the RWA can not
describe the essential feature of the evolution of  the entanglement
beyond the weak coupling regime and is therefore broken down under strong coupling.

Importantly, this physical condition can be realized in the recent
strong coupling experiments ~\cite{niemczyk1}. In the non-Markovian
limit ($\gamma/\lambda{\ll}1$), the spectral function of baths is
reduced to the single mode case by
$J(\omega){\approx}\frac{\lambda\gamma}{2}\delta(\omega-\omega_0)$.
Hence, the system can be approximately simplified to two independent
Rabi models. The corresponding effective system-bath coupling
constant is given by $g=\sqrt{\frac{\lambda\gamma}{2}}$, so
$g/\omega_0=0.11$ for $f=0.1$,  which can be practically realized in
the recent experiments of the superconducting qubits coupled to LC
resonators, where the coupling constant $g/\omega_0$ already reached
to ten percentages~\cite{niemczyk1}.

In the Markovian regime, the  entanglement vanishes quickly and
permanently beyond the weak coupling as shown in
Fig.~\ref{TATBcompare} (b) and~\ref{TATBcompare}(c), mainly due to
the generation of extra phonons with increasing coupling, which
disturb the pairwise entanglement.  No revival of the entanglement
appears in this case.  Interestingly,  the evolution behavior is
almost  unchanged as the increasing coupling strength for
$\lambda\ge 0.5$, which follows that $\lambda=0.5$ is strong enough
in this case. However, in the non-Markovian regime,  the
entanglement dynamics show essential different behavior with the
system-bath coupling. The revival of the entanglement after the ESD
occurs at $\lambda=0.5$,  but this revival phenomena never happens in
the strong coupling regime, as shown  in Fig.~\ref{TATBcompare}(c)
for $\lambda=2$. We propose that  the more extra phonons  activated
by the CRT would suppress the feedback from the bath to the
qubit~\cite{chen1} even in the non-Markovian regime, resulting in
the permanent disentanglement. The effective system-bath coupling
strength for $f=0.1$ in Fig.~\ref{TATBcompare}(c) can be estimated
as  $g/\omega_0=0.45$, which is hopefully realized in the near
future~\cite{solano1,solano2}.

\begin{figure}[tbp]
\includegraphics[scale=0.5]{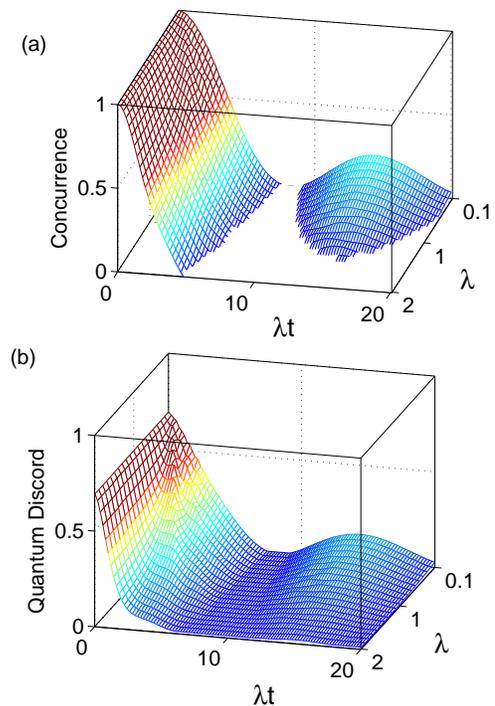}
\caption{Effect of system bath interaction on dynamics of quantum
correlation of two qubits of two independent baths  with initial
anti-correlated  Bell state  $|\Phi(\frac{1}{\sqrt{2}}){\rangle}$.
$\omega_0=1$, $\gamma=0.1\lambda$.} \label{TATBcmp3d}
\end{figure}

Then we  investigate the dynamics of the  QD for two qubits in two
independent baths to explore the influence of the CRT. The QD  in
the RWA has been discussed  by Wang \emph{et al.}~\cite{wang1} and
Fanchini \emph{et al.}~\cite{fanchini1}. In the lower panel of
Fig.~\ref{TATBcompare}(d), we display the evolution of the  QD  for
the same parameters as the upper panel by both the hierarchy
approach and RWA. In the weak coupling regime, it behaves similar to
the concurrence in Fig.~\ref{TATBcompare}(a)  for both the Markovian
and non-Markovian evolutions, where the CRT can be ignored. Beyond
the weak coupling regime, as shown in Fig.~\ref{TATBcompare}(e)
and~\ref{TATBcompare}(f), the CRT drives the
 QD   to deviate from those in RWA obviously. Specifically
for the non-Markovian case, the  QD  decays dramatically and
shows weak revival signal, which mainly attributes to the excitation
of the phonons in baths. For the Markovian case, the  QD
decreases faster than that in RWA and decay to zero  asymptotically.

To see the comprehensive effect of the system-bath interaction, we
plot the dynamics of the quantum correlation as a function of both
$t$ and $\lambda$ in Fig.~\ref{TATBcmp3d} for initial
anti-correlated Bell state. Both the entanglement and the discord
are suppressed monotonically with the increasing interaction. The
revival phenomena for entanglement is not present completely as the
coupling strength exceeds a critical value, e.g. $\lambda_c$ is
about $1.5$ for the parameters in Fig.~\ref{TATBcmp3d} (a). The
region of  ESD becomes wider with the increase of the system-bath
coupling. But Fig.~\ref{TATBcmp3d} (b) reveals that the
 QD  never vanishes suddenly, consistent with the previous observation
concluded from an arbitrary Markovian evolution ~\cite{ferraro1}. Therefore,
such a limitation is removed in the present observation
in the exact evolution which definitely includes the non-
Markovian effect. No sudden death of the QD  is universal in this sense,
in sharp contrast with the entanglement.
More over, in our exact solution for the full model without the RWA,
we actually have not found the real zero QD at any time, despite arbitrary small.
This nonzero nature for the QD maybe intrinsic at least for the spin-boson model.
Previous observation in the RWA that the QD can vanish at some discreet times~\cite{wang1,fanchini1}
may be an artificial result due to the absence of the CRT.
\begin{figure}[tbp]
\includegraphics[scale=0.4]{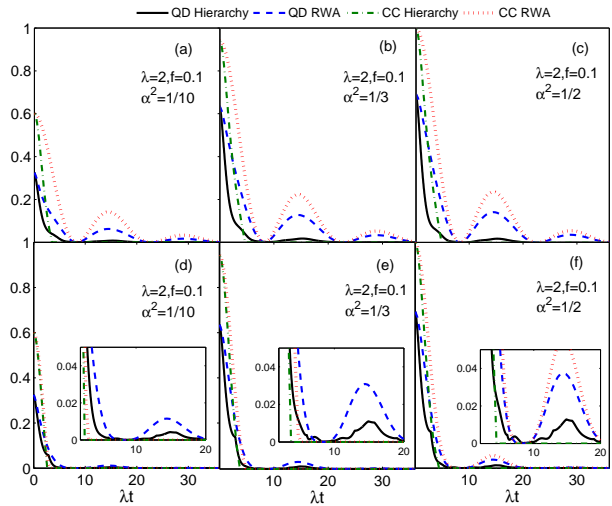}
\caption{Non-Markovian dynamics of quantum correlation of two qubits
of two independent baths with  initial anti-correlated Bell states
$|\Phi(\alpha){\rangle}$ (upper panel)  and correlated Bell states
$|\Psi(\alpha){\rangle}$ (low panel). The insets in low panel are
the enlarged views. $\omega_0=1$, $\lambda=2$, and
$\gamma=0.1\lambda$. } \label{TATBat}
\end{figure}

The essential feature  for the dynamics of the quantum entanglement
and discord under the RWA  is not altered qualitatively with the
system-bath coupling in the same non-Markovian  or  Markovian
regime. However, it is not that case in the real exact solution
without the RWA. For the weak system-bath interaction, we have shown
above that the RWA can basically give the right description of the
quantum correlation of two separate spin-boson models  for both
Markovian and non-Markovian regimes. While in the strong coupling
regime, the CRT should be necessarily included for correctly
depicting the phonons generation from the baths. And the quantum
correlation shows significantly different behavior  from that in
RWA, especially in non-Markovian case. Hence, we focus on the
dynamics of the quantum correlation in the strong coupling and
non-Markovian regime in the following with various mostly used
initial states.

\subsection{Non-Markovian dynamics at strong coupling}

\begin{figure}[tbp]
\includegraphics[scale=0.4]{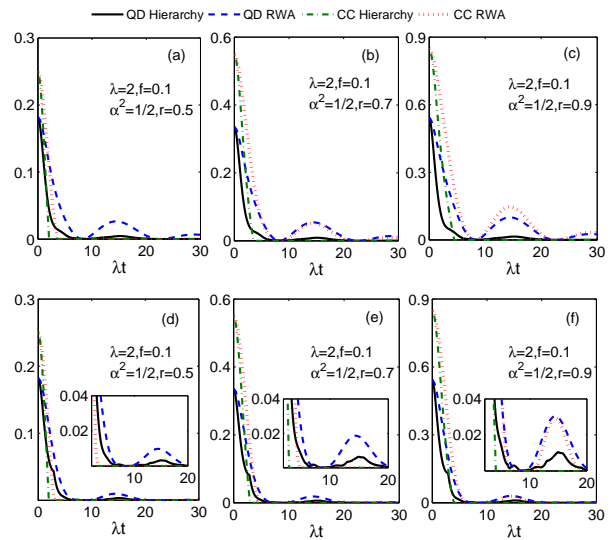}
\caption{Non-Markovian dynamics of quantum correlation of two qubits
of the two independent baths with extended Werner like initial state
$\rho_{AB}(0)=r|\xi{\rangle}{\langle}\xi|+\frac{1-r}{4}\mathbb{I}$:
$|\xi{\rangle}=|\Phi(\frac{1}{\sqrt{2}}){\rangle}$ (upper panel) and
$|\xi{\rangle}=|\Psi(\frac{1}{\sqrt{2}}){\rangle}$ (lower panel) .
The insets in the lower panel are the corresponding enlarged views.
$\omega_0=1$, $\lambda=2$, and $\gamma=0.1\lambda$. }
\label{TATBratio}
\end{figure}

First, we examine the dependence of the dynamics of the quantum
correlation on the   value of $\alpha$ for the initial  Bell state
in Eq.~(\ref{tisp61}), which characterize  weights of the two
superposed anti-correlated spin states. Since the quantum
correlation is similar in the regime $\alpha^2{\in}(0,1/2)$ to that
in $(1/2,1)$, the first zone is selected to study the correlation
evolution in Fig.~\ref{TATBat}(a)-\ref{TATBat}(c). In RWA, both the
concurrence and the  QD  show damping oscillations with stronger
amplitude as $\alpha^2$ increases. The other evident feature is that
they only vanish at some discrete moments.  When the CRT is
included, the concurrence exhibits ESD after finite time evolution
in the whole regime. The starting time of ESD is much earlier than
first vanishing moment in the RWA. The QD first decreases  almost to
zero and then revives with amplitude  much too weaker than  that in
the  RWA, mainly attributed to the suppress from the phonons.

Fig.~\ref{TATBat}(d)-\ref{TATBat}(f) presents the results for the
another initial Bell state in Eq.~(\ref{tisp62}).  It is known that
the discord dynamics for $\alpha^2{\in}(0,1)$ is similar to the
concurrence for $\alpha^{2}{\in}(1/2,1)$. Hence, we also select
these initial states for comparison as in Ref.~\cite{fanchini1}. The
exact dynamics shows that the ESD occurs very quickly and irrelevant
with the value of $\alpha$. The QD first decreases almost to zero,
and then revives weakly. The revival  amplitude becomes stronger
with the value of $\alpha$. For the same $\alpha$, the revival
amplitude is much too weaker than that in the RWA, due to the CRT
effect, as demonstrated in the inset.

Finally, we assume the extended Werner like state
($\rho_{AB}(0)=r|\xi{\rangle}{\langle}{\xi}|+\frac{1-r}{4}\mathbb{I}$)
as the initial state and see what happen to the evolution of these
quantum correlations. The main results are list in
Fig.~\ref{TATBratio}. As one classical kind of the mixed states, it
is important in quantum information
processing~\cite{werner1,popescu1,hagley1}. As the initial state
with $|\xi{\rangle}=|\Phi(\frac{1}{\sqrt{2}}){\rangle}$, the
concurrence in RWA shows complete death in $r{\in}(0,1/2)$, whereas
the discord in the regime $r{\in}(0,1)$ behaves similar with the
concurrence in $r{\in}(1/2,1)$~\cite{wang1}. Hence, we study the
quantum correlation in the regime $r{\in}(1/2,1)$ to explore the
effect of the CRT in Fig.~\ref{TATBratio}(a)-\ref{TATBratio}(c). The
entanglement vanishes suddenly and permanently for all values of
$r$, RWA ones  show damping oscillations for  large $r$ (e.g.
$r=0.7$, $0.9$). The QD is also strongly suppressed first, and then
revives weakly. The revival amplitude increase slightly with $r$. It
is again shown that the QD is not so fragile as the entanglement. If
$|\Psi(\frac{1}{\sqrt{2}}){\rangle}$ is  replaced by $|\xi{\rangle}$
in the initial extended Werner like state,  no essential different
behavior can be observed except that the revival amplitude becomes
much too weaker, as demonstrated in
Fig.~\ref{TATBratio}(d)-\ref{TATBratio}(f).

For the three kinds of initial states studied above, it is generally
find that the larger initial quantum correlations is, the stronger
they evolves at the same moments.

\section{two qubits coupled to one common bosonic bath}

The Hamiltonian of two qubits coupled to the common bath
$H=H_s+H_b+H_{sb}$ reads
\begin{eqnarray}~\label{tqob1}
H_s&=&\frac{\omega_0}{2}\sum_{a=A,B}\sigma^a_z, \nonumber\\
H_b&=&\sum_k{\omega_k}b^{\dag}_kb_k, \nonumber\\
H_{sb}&=&V\sum_k(g_kb_k+g^*_kb^{\dag}_k),
\end{eqnarray}
where the notations are the same as those in Eq. (2).
If the spectrum of the common bath is also
Lorentz distribution
$J(\omega)=\frac{1}{2\pi}\frac{\lambda\gamma^2}{(\omega-\omega_0)^2+\gamma^2}$,
and the correlation function of the bath should be also
$C(t)=\frac{\lambda\gamma}{2}e^{-(\gamma+i\omega_0)t}$. By choosing
the proper auxiliary influence functional in Appendix, the hierarchy
equation is given by~(\ref{app21})~\cite{ma1}
\begin{eqnarray}~\label{tqob2}
\frac{{\partial}\rho_{\vec{n}}(t)}{{\partial}t}&=&-(iH^{\times}_S+\vec{n}\cdot\vec{\nu})\rho_{\vec{n}}(t)-i\sum_{k=1,2}V^{\times}\rho_{\vec{n}+\vec{e}_k}(t) \nonumber\\
&&-\frac{i\gamma\lambda}{4}\sum_{k=1,2}n_k[V^{\times}+(-1)^kV^{o}]\rho_{\vec{n}-\vec{e}_k}(t), \nonumber\\
\end{eqnarray}
where $\vec{n}=(n_1,n_2)$, the unit vector $\vec{e}_1=(1,0)$ and
$\vec{e}_2=(0,1)$. The reduced  density matrix for qubits is
$\rho_{AB}(t)=\rho_{(0,0)}(t)$, by which  we can derive the quantum
correlation numerically.

In RWA, the system-bath interaction becomes
$H^R_{sb}=(\sigma^A_++\sigma^B_+)\sum_kg_kb_k+(\sigma^A_-+\sigma^B_-)\sum_kg^*_kb^{\dag}_k$.
If we represent the Hamiltonian in the basis
$\{|0{\rangle}=|00{\rangle},|+{\rangle}=\frac{1}{2}(|10{\rangle}+|01{\rangle}),
|-{\rangle}=\frac{1}{\sqrt{2}}(|10{\rangle}-|01{\rangle}),|2{\rangle}=|11{\rangle}\}$,
$|-{\rangle}$ is found to be isolate from others. Hence we can
rewrite the Hamiltonian in three level form\cite{garraway4}
\begin{eqnarray}~\label{tqob3}
H_0&=&2\omega_0|2{\rangle}{\langle}2|+\omega_0|+{\rangle}{\langle}+|-\omega_0, \\
H_{int}&=&\sqrt{2}\sum_kg_ka_k(|+{\rangle}{\langle}0|+|2{\rangle}{\langle}+|)+H.c. \nonumber
\end{eqnarray}
For the single particle excitation or double excitation case, the
motion equation of the system density matrix can be derived by the
called pseudomode mater equation in the interacting
picture~\cite{garraway1,garraway2,garraway3,garraway4,mazzola1,fanchini1}
\begin{eqnarray}~\label{tqob4}
\frac{{\partial}\rho_I(t)}{{\partial}t}&=&-i[V,\rho_I(t)] \nonumber\\
&&-\gamma(a^{\dag}a\rho_I(t)+\rho_I(t)a^{\dag}a-2a\rho_I(t)a^{\dag}),
\end{eqnarray}
where $a^{\dag}~(a)$ is the creator (annihilator) of the bosonic pseudomode,
\begin{eqnarray}~\label{tqob5}
V=\sqrt{\lambda\gamma}(a|+{\rangle}{\langle}0|+a^{\dag}|0{\rangle}{\langle}+|+a|2{\rangle}{\langle}+|+a^{\dag}|+{\rangle}{\langle}2|). \nonumber
\end{eqnarray}
and $\rho_I(t)=e^{iH_0t}\rho(t)e^{-iH_0t}$. By integrating the
bosonic part of the density matrix, the system reduced density is
given by $\rho_{AB}(t)=\textrm{Tr}_a\{\rho(t)\}$. For single
excitation, an alternatively exact solution has also been done in
Ref.~\cite{maniscalco1}. If the initial state is chosen by $X$ form
in Eq.~(\ref{tisp4}), the concurrence is derived by~\cite{mazzola1}
\begin{eqnarray}~\label{cc}
C(t)&=&2\max\{0,|\rho_{23}|-\sqrt{\rho_{11}\rho_{44}},|\rho_{14}|-\sqrt{\rho_{22}\rho_{33}}\}.\nonumber\\
\end{eqnarray}

For the common bath, the entanglement evolution initiated from the
anti-correlated Bell state without the RWA has been studied by Ma
\emph{et al.}~\cite{ma1}. On the other hand, in the previous studies
of the dynamics of  QD under the RWA, the correlated Bell state  is
usually selected as the initial
state~\cite{wang1,fanchini1,garraway4,mazzola1}. Therefore in the
present exact study without the RWA, we will focus on the initial
correlated Bell state.

\begin{figure}[tbp]
\includegraphics[scale=0.4]{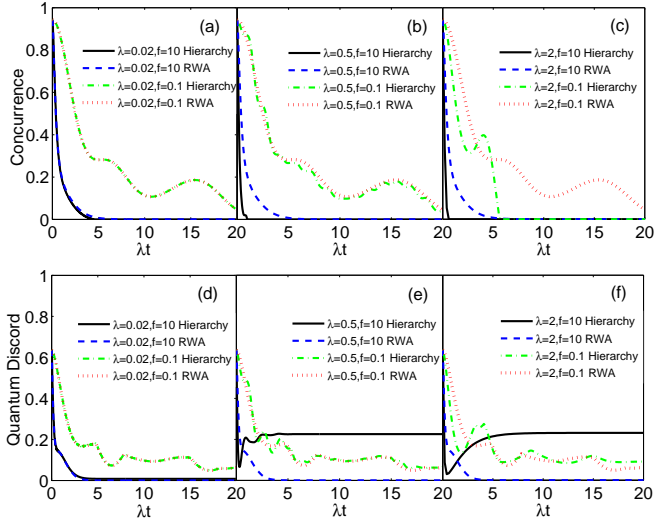}
\caption{Dynamics of quantum correlation of two qubits coupled to
the common  bath  with initial correlated Bell state:
$|\Psi(\frac{1}{\sqrt{3}}){\rangle}$. $\omega_0=1$,
$\gamma=f\lambda$. } \label{comcompare}
\end{figure}

First, we exhibit  the dynamics of the pairwise concurrence of the
two qubits coupled to a common  bath with and without the RWA
initiated from the Bell state in Eq.~$(\ref{tisp61})$ with
$\alpha=1/\sqrt{3}$   in the upper panel of Fig.~\ref{comcompare}
for various coupling strength  in  both Markovian and non-Markovian
regimes. Similarly, for the weak coupling ($\lambda=0.02$), the CRTs
take little effect for both dynamics in different evolution regimes,
showing the same results as in Ref.~\cite{fanchini1}.  In the
intermediate coupling regime ($\lambda=0.5$), the ESD occurs more
quickly than that in the RWA in the Markovian regime, and the
evolution  shows slight deviations from  the RWA ones with damping
oscillations in the non-Markovian regime.  When  the strong coupling
regime is reached, the CRT takes remarkable effects in both
evolution regimes, where concurrence vanishes completely and
permanently without the RWA. Hence, the CRT also suppress the
entanglement dramatically as the system-bath coupling for the common
bath, similar to those for the independent baths.

Then we turn to the relevant  QD, as shown in the lower panel of
Fig.~\ref{comcompare}.  It is interesting to note that in the
non-Markovian regime,   the CRTs take  very limited effect with the
increasing coupling. Even at the strong coupling, only slightly
deviations are observed.  More interestingly, the QD takes finite
value at any time, and take a moderate finite value asymptotically
in the long time limit, for arbitrary coupling strength. In the
Markovian regime, the QD tends to a remarkable value in the
intermediate and strong coupling regime. Actually, even at weak
coupling like $\lambda=0.05$, the asymptotical value of QD takes a
considerable non-zero value.

From the numerical iteration of the hierarchy equation in
Eq.~(\ref{tqob2}) with Bell correlated initial state under Markov
regime (e.g. $f=10$ in Fig.~\ref{comcompare}), we find that the form
of the two qubits reduced density matrix under the same basis as in
Eq.~(\ref{tisp4}) always to be $X$ type as
\begin{eqnarray}
\rho(t)=
\begin{pmatrix}
a(t) & 0 & 0 & w(t)\\
0 & b(t) & b(t) & 0\\
0 & b(t) & b(t) & 0\\
w^{*}(t) & 0 & 0 & d(t)
\end{pmatrix}.
\end{eqnarray}
In the weak system-bath coupling (say $\lambda=0.02$), $a(t)$ and
$b(t)$ decrease gradually to $0$ after the finite time evolution,
whereas $d(t)$ rises to reach $1$. Hence both concurrence and the
discord decay to $0$ exhibit monotonically as indicated in
Fig.~\ref{comcompare}(a) and~\ref{comcompare}(d), similar to that in
the RWA. When the coupling strength becomes stronger, e.g.
$\lambda=0.5, 2.0$, the phonons are excited considerably due to the
strong coupling~\cite{chen1}. According to the CRT
$\sum_kg^{*}_kb^{\dag}_k(\sigma^A_++\sigma^B_+)$, the qubits are
also excited accompanying the extra excitations of the phonons,
which stabilize $a(t)$ and $b(t)$ in long time evolution. It is
interesting to observe that they tend to $1/3$ and $1/6$  in the
strong coupling limit at $t{\rightarrow}{\infty}$. The corresponding
reduced density matrix is then described by
$\rho(t{\rightarrow}\infty)=\frac{1}{3}(|00{\rangle}{\langle}00|+|11{\rangle}{\langle}11|+|+{\rangle}{\langle}+|)$.
which is not altered by different $\alpha^2$. Under this separable
mixed state, the disentanglement occurs naturally from
Eq.~(\ref{cc}), and the QD is stabilized at $0.23$.

\begin{figure}[tbp]
\includegraphics[scale=0.4]{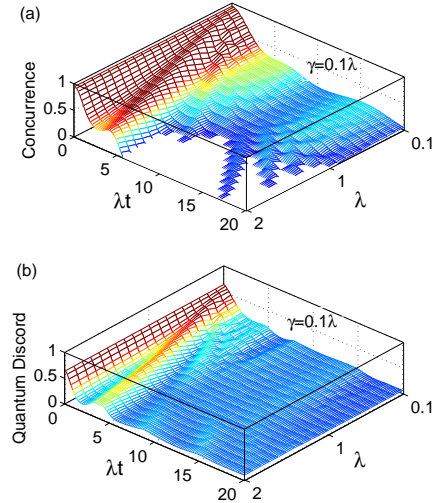}
\caption{Dynamics of quantum correlation of two qubits coupled to
the common  bath with initial correlated Bell state $|\Psi(\frac{1}{\sqrt{3}}){\rangle}$ (a) concurrence and
 (b) QD.  $\omega_0=1$, $\gamma=0.1 \lambda$. } \label{comon3D}
\end{figure}

To see the overall effect of the system-bath interaction on the
dynamics of the quantum correlation beyond the weak coupling regime,
we draw the corresponding $3$D plots for $\lambda \ge 0.1$  in
Fig.~\ref{comon3D} in the non-Markovian regime.  For the
entanglement, one can find that there are two regimes. One is
damping oscillation at $\lambda<1$. The other is the regime where
the entanglement revives   following the ESD, and the revival will
be suppressed and finally disappear with the further increase of the
coupling regime. Very interestingly, after weak oscillations, the QD
is generally robust in the long time evolution  in arbitrary
coupling regime, which is very useful as the quantum information
resource.

\section{conclusion}

In summary, without performing any approximation, we exactly
investigate quantum correlations in two typical kinds of spin-boson
models by applying the hierarchy approach. The corresponding RWA
results are also reproduced  for comparison. The Markovian and
non-Markovian effects of two kinds baths in strong system-bath
interaction on the dynamics of the quantum correlation are revealed
in detail, which demonstrates that the CRT should be necessarily
included beyond the weak coupling regime to  capture the essential
features of quantum correlations correctly.

For two independent baths, the CRT  suppresses both the entanglement
and the QD dramatically, due to the activations of phonons in baths
as the system-bath coupling increases. If initiated from the anti-correlated
Bell state, The ESD is driven by extra phonons activated with the
increasing system-bath coupling, which is absent in the RWA study.
It follows that previous picture under the RWA is essentially
modified with the increasing coupling. The QD is found to be always
higher than zero, despite extremely small, also different from that
in the RWA. The dynamics with other initial conditions is also
discussed at the strong coupling and non-Markovian regime. The
general trend  is that the larger initial quantum correlations, the
stronger they evolve.

For the model of two qubits coupled to the common bath, the
suppression of the entanglement by the CRT is also observed. It is
interesting to find that  the non-classical correlation reserved by
the QD can be enhanced as the coupling strength, in contrast with
the entanglement. It follows that while the entanglement becomes
fragile with the increasing coupling to the surrounding environment,
the QD is however robust in the long time evolution. In recently
experiments, the strong system-bath coupling has been reached in
circuit QED systems~\cite{niemczyk1}, where $g/\omega_0{\approx}0.1$
for multi modes and the RWA description is broken down. The present
study for the evolution of the quantum correlations beyond the RWA
may motivate the corresponding experimental studies  based on these
strong coupling systems.

\section{Acknowledgement}

This work was supported by National Natural Science Foundation of
China under Grant No.~11174254 and the National Basic Research
Program of China under Grant No. 2009CB929104.

\appendix

\section{basic concept and application of the Hierarchy equation }
The Hamiltonian of the system-bath coupling system is modeled
as~$H=H_s+H_b+H_{sb}$~. Then the time dependent density matrix is given by
$\rho(t)=e^{-iHt}\rho(0)e^{iHt}$. In the interacting picture, the
corresponding density matrix reads
$\rho_I(t)=U_I(t)\rho(0)U^{\dag}_I(t)$, with
$U_I(t)=e^{-i\int^t_0d{\tau}H_{sb}(\tau)}$. Under the specific
representation $\{|x,\alpha{\rangle}\}$,
\begin{eqnarray}~\label{app1}
&&{\langle}x_t,\alpha_t|\rho_I(t)|x^{'}_t,\alpha^{'}_t{\rangle} \nonumber\\
&=&{\langle}x_t,\alpha_t|U_I(t)\rho(0)U^{\dag}_I(t)|x^{'}_t,\alpha^{'}_t{\rangle} \nonumber\\
&=&\int{dx_0}d{x^{'}_0}\int{d\alpha_0}d{\alpha^{'}_0}{\langle}x_t,\alpha_t|U_I(t)|x_0,\alpha_0{\rangle} \nonumber\\
&{\times}&{\langle}x_0,\alpha_0|\rho(0)|x^{'}_0,\alpha^{'}_0{\rangle}{\langle}x^{'}_0\alpha^{'}_0|U^{\dag}_I(t)|x^{'}_t,\alpha^{'}_t{\rangle}.
\end{eqnarray}
It is known that $\rho^I_s(t)=\textrm{Tr}_b\{\rho_I(t)\}$. Hence by integrating the bath degree of freedom, the reduced system density is
\begin{eqnarray}~\label{app2}
&&{\langle}a_t|\rho^I_S(t)|\alpha^{'}_t{\rangle} \nonumber\\
&=&\int{dx_t}{\langle}x_t,\alpha_t|\rho_I(t)|x_t,\alpha^{'}_t{\rangle} \nonumber\\
&=&\int{dx_t}{dx_0}{dx^{'}_0}\int{d\alpha_0}{d\alpha^{'}_0}
{\langle}x_t,\alpha_t|U_I(t)|x_0,\alpha_0{\rangle} \nonumber\\
&{\times}&{\langle}x^{'}_0,\alpha^{'}_0|U^{\dag}_I(t)|x_t,\alpha^{'}_t{\rangle}
{\langle}x_0,\alpha_0|\rho(0)|x^{'}_0,\alpha^{'}_0{\rangle}.
\end{eqnarray}
Considering the propagator in functional form
\begin{eqnarray}~\label{app3}
{\langle}x_t,\alpha_t|U_I(t)|x_0,\alpha_0{\rangle}=\int^{x_t}_{x_0}\mathcal{D}[x(\tau)]\int^{\alpha_t}_{\alpha_0}\mathcal{D}[\alpha(\tau)]e^{iS[x(\tau),\alpha(\tau)]} \nonumber
\end{eqnarray}
and the initial factorized state $\rho(0)=\rho_s(0){\otimes}\rho_b$~($\rho_b=\exp(-{\beta}H_b)/Z$ is the equilibrium state),
the reduced density matrix is reexpressed by
\begin{eqnarray}~\label{app4}
&&{\langle}{\alpha_t}|\rho^{I}_S(t)|\alpha^{'}_t{\rangle} \\
&=&
\int{d\alpha_0}{d\alpha^{'}_0}{\langle}{\alpha_0}|\rho_S(t)|\alpha^{'}_0{\rangle}  \nonumber\\
&{\times}&\int^{\alpha_t}_{\alpha_0}\mathcal{D}[\alpha(\tau)]
\int^{\alpha^{'}_t}_{\alpha^{'}_0}\mathcal{D}[\alpha^{'}(\tau)]\int{dx_t}{dx_0}{dx^{'}_0}{\langle}x_0|\rho_B|x^{'}_0{\rangle} \nonumber\\
&{\times}&\int^{x_t}_{x_0}\mathcal{D}[x(\tau)]\int^{x_t}_{x^{'}_0}\mathcal{D}^*[x^{'}(\tau)]e^{iS[x(\tau),\alpha(\tau)]-iS[x^{'}(\tau),\alpha^{'}(\tau)]}.\nonumber
\end{eqnarray}
Then the functional field is introduced by
\begin{eqnarray}~\label{app5}
&&\mathcal{F}[\alpha(\tau),\alpha^{'}(\tau)] \nonumber\\
&=&\int{dx_t}{dx_0}{dx^{'}_0}{\langle}x_0|\rho_b|x^{'}_0{\rangle} \nonumber\\
&{\times}&\int^{x_t}_{x_0}\mathcal{D}[x(\tau)]\int^{x_t}_{x^{'}_0}\mathcal{D}^*[x^{'}(\tau)]e^{iS[x(\tau),\alpha(\tau)]-iS[x^{'}(\tau),\alpha^{'}(\tau)]} \nonumber\\
&=&\textrm{Tr}_b\{\rho_bU^{\dag}_b[\alpha^{'}]U_b[\alpha]\},
\end{eqnarray}
where $\langle{x_t}|U_b[\alpha(\tau)]|x_0{\rangle}=\int^{x_t}_{x_0}\mathcal{D}[x(\tau)]e^{iS[x(\tau),\alpha(\tau)]}$.
The functional field generally satisfies $\mathcal{F}[\alpha,\alpha^{'}]{\le}1$, where $\mathcal{F}[\alpha,\alpha^{'}]=1$ if there is no system-bath interaction.

Here, we include the harmonic bath with the Hamiltonian as
$H_b=\sum_k\frac{\hat{p}^2_k}{2m_k}+\frac{m_k\Omega^2_k}{2}\hat{x}^2_k$.
Without loss of generality, we firstly consider only one mode case as
$H_b=\frac{\hat{p}^2}{2m}+\frac{m\Omega^2}{2}\hat{x}^2$, and the
system-bath coupling term is  $H_{sb}=\hat{Q}\hat{x}$. In the
interaction picture, $H_{sb}=\hat{Q}(t)\hat{x}(t)$, and
$U^I_b[\alpha(\tau)]=e^{-i\int^{t}_0{d\tau}Q[\alpha(\tau)]\hat{x}(\tau)}$.
Through the tedious by standard Feynman-Vernon  influence functional
calculation, we derive the influence functional as
$\mathcal{F}[\alpha,\alpha^{'}]=e^{-\Phi[\alpha,\alpha^{'}]}$, where
the influence phase is
\begin{eqnarray}~\label{app6}
\Phi[\alpha,\alpha^{'}]&=&\int^t_0{dt^{'}}\int^{t^{'}}_0ds\{Q[\alpha(t^{'})]-Q[\alpha^{'}(t^{'})]\} \\
&{\times}&\{C(t^{'}-s)Q[\alpha(s)]-C^*(t^{'}-s)Q[\alpha^{'}(s)]\}, \nonumber
\end{eqnarray}
with corresponding bath correlation function
\begin{eqnarray}~\label{app7}
C(\tau)=\frac{1}{m\Omega}[\coth({\beta\Omega}/2)\cos{\Omega\tau}-i\sin{\Omega\tau}].
\end{eqnarray}
Here we should note that the variable $Q[\alpha]$ is a real number.
Then we generalize the results to the multi-mode case, where
the system-bath interaction now is $H_{sb}=\sum_a\hat{Q}_a\hat{F}_a$, and $\hat{F}_a=\sum_kc^a_k\hat{x}^a_k$.
The  influence functional becomes
\begin{eqnarray}~\label{app8}
\mathcal{F}[\alpha(\tau),\alpha^{'}(\tau)]&=&\exp(-\sum_a\int^{t}_0{d\tau}
\{Q_a[\alpha(\tau)]-Q_a[\alpha^{'}(\tau)]\} \nonumber\\
&&{\times}\{\tilde{Q}_a[\alpha(\tau)]-\tilde{Q}^{*}_a[\alpha^{'}(\tau)]\}),
\end{eqnarray}
where
\begin{eqnarray}~\label{app9}
\tilde{Q}_a[\alpha(t)]=\sum_{a'}\int^{t}_0{d\tau}C_{a{a'}}(t-\tau)Q_{a'}[\alpha(\tau)].
\end{eqnarray}
The correlation function in independent baths is
\begin{eqnarray}~\label{app10}
C_{a{a'}}(\tau)&=&{\langle}F_a(\tau)F_{a'}(0){\rangle} \\
&=&\sum_k\frac{\delta_{a{a'}}{c^{a}_k}^2}{2m^a_k\Omega^a_k}[\coth(\frac{\beta_a\Omega^a_k}{2})\cos(\Omega^a_k\tau)-i\sin(\Omega^a_k\tau)]. \nonumber
\end{eqnarray}
By using  the spectral function $J_a(\omega)=\frac{1}{2}\sum_k\frac{{c^{a}_k}^2}{m^a_k\Omega^a_k}\delta(\omega-\Omega^a_k)$,
we finally have
\begin{eqnarray}~\label{app11}
C_{a{a'}}(t)=\delta_{a{a'}}\int{d\omega}J_a(\omega)[\coth(\frac{\beta_a\omega}{2})\cos(\omega\tau)-i\sin(\omega\tau)]. \nonumber
\end{eqnarray}

In the schrodinger picture, the motion equation of the system density is
\begin{eqnarray}~\label{app12}
\frac{{\partial}\rho_S(t)}{{\partial}t}=-i[H_S,\rho_S(t)]+e^{-iH_St}\frac{{\partial}\rho^I_S(t)}{{\partial}t}e^{iH_St}.
\end{eqnarray}
From Eq.~(\ref{app4}) and (\ref{app5}), we have
\begin{eqnarray}~\label{app13}
\frac{{\partial}\rho^I_S(t)}{{\partial}t}&=&\int{d\alpha_0}{d\alpha^{'}_0}{\langle}\alpha_0|\rho_S(0)|\alpha^{'}_0{\rangle} \\
&{\times}&\int^{\alpha_t}_{\alpha_0}\mathcal{D}[\alpha(\tau)]\int^{\alpha^{'}_t}_{\alpha^{'}_0}\mathcal{D}^{*}[\alpha^{'}(\tau)]
\frac{{\partial}\mathcal{F}[\alpha,\alpha^{'}]}{{\partial}t}.\nonumber
\end{eqnarray}
The key step here is to connect time differentiation of the density matrix with that of the influence functional.
The proper way is employing the hierarchy equation by introducing the auxiliary influence functional
\begin{eqnarray}~\label{app14}
\rho^{I}_S(n,t)&=&\int{d\alpha_0}{d\alpha^{'}_0}{\langle}\alpha_0|\rho_S(0)|\alpha^{'}_0{\rangle} \\
&{\times}&\int^{\alpha_t}_{\alpha_0}\mathcal{D}[\alpha(\tau)]
\int^{\alpha^{'}_t}_{\alpha^{'}_0}\mathcal{D}^{*}[\alpha^{'}(\tau)]\mathcal{F}_n[\alpha,\alpha^{'}]. \nonumber
\end{eqnarray}
While such auxiliary choice is closely dependent on the specific
form of the bath correlation function. The hierarchy equation used in the
present paper is described in the following.

For two independent baths   with Lorentz type
spectral function
$J_a(\omega)=\frac{1}{2\pi}\frac{\lambda_a\gamma^2_a}{(\omega-\omega^a_0)^2+\gamma^2_a}$,
the correlation function simply is
\begin{eqnarray}~\label{app15}
C_{aa}(\omega)=\frac{\lambda_a\gamma_a}{2}\exp[-(\gamma_a+i\omega^a_0)t].
\end{eqnarray}
according to the following formula
\begin{eqnarray} \label{app16}
\frac{\partial}{{\partial}t}\tilde{Q}_a[\alpha(t)]=\frac{\lambda_a\gamma_a}{2}Q_a[\alpha(t)]-(\gamma_a+i\omega^a_0)\tilde{Q}_a[\alpha(t)], \nonumber
\end{eqnarray}
we introduce the auxiliary  influence functional of each subsystem as
\begin{eqnarray}  \label{app17}
\mathcal{F}_{a,\vec{k}}[\alpha,\alpha^{'}]=(\tilde{Q}_a[\alpha(t)])^{k_1}(\tilde{Q}^{*}_a[\alpha^{'}(t)])^{k_2}\mathcal{F}[\alpha,\alpha^{'}],
\end{eqnarray}
where $Q_a=\sigma^a_x,~(a=A,B)$ and index $\vec{k}=(k_1,k_2)$. Then the whole influence
functional  becomes
\begin{eqnarray}  \label{app18}
\mathcal{F}_{\vec{n},\vec{m}}[\alpha,\alpha^{'}]=\mathcal{F}_{A,\vec{n}}[\alpha,\alpha^{'}]\mathcal{F}_{B,\vec{m}}[\alpha,\alpha^{'}].
\end{eqnarray}
Finally, the evolution of the density matrix reads
\begin{eqnarray} \label{app19}
\frac{\partial}{{\partial}t}\rho_{\vec{n},\vec{m}}(t)&=&-(iH^{\times}_S+\vec{n}\cdot\vec{\nu}_{A}+\vec{m}\cdot\vec{\nu}_{B})\rho_{\vec{n},\vec{m}}(t) \\
&&+\sum_{k=1,2}(-1)^{k}Q^{\times}_A\rho_{\vec{n}+\vec{e}_k,\vec{m}}(t) \nonumber\\
&&+\frac{\lambda_A\gamma_A}{4}\sum_{k=1,2}n_k[Q^{o}_{A}+(-1)^{k+1}Q^{\times}_{A}]\rho_{\vec{n}-\vec{e}_{k},\vec{m}}(t)\nonumber\\
&&+\sum_{k=1,2}(-1)^kQ^{\times}_B\rho_{\vec{n},\vec{m}+\vec{e}_k}(t) \nonumber\\
&&+\frac{\lambda_B\gamma_B}{4}\sum_{k=1,2}m_k[Q^{o}_B+(-1)^{k+1}Q^{\times}_B]\rho_{\vec{n},\vec{m}-\vec{e}_k}(t),\nonumber
\end{eqnarray}
where unit vector $\vec{e}_1=(1,0)$ and $\vec{e}_2=(0,1)$ $\vec{\nu}_a=(\nu^a_+,\nu^a_-)$, with $\nu^a_+=\gamma_a+i\omega^a_0$
and $\nu^a_-=\gamma_a-i\omega^a_0$. $A^{\times}B=AB-BA$ and $A^{o}B=AB+BA$.

For common bosonic bath with the similar Lorentz distribution, if we choose the auxiliary  influence functional
\begin{eqnarray} \label{app20}
\mathcal{F}_{\vec{n}}[\alpha,\alpha^{'}]=i^{n_1}(\tilde{Q}^{*}[\alpha^{'}(t)])^{n_1}(-i)^{n_2}(\tilde{Q}[\alpha(t)])^{n_2}\mathcal{F}[\alpha,\alpha^{'}], \nonumber
\end{eqnarray}
the hierarchy equation can be  expressed by
\begin{eqnarray} \label{app21}
\frac{{\partial}\rho_{\vec{n}}(t)}{{\partial}t}&=&-(iH^{\times}_S+\vec{n}\cdot\vec{\nu})\rho_{\vec{n}}(t)-i\sum_{k=1,2}Q^{\times}\rho_{\vec{n}+\vec{e}_k}(t) \nonumber\\
&&-\frac{i\gamma\lambda}{4}\sum_{k=1,2}n_k[Q^{\times}+(-1)^kQ^{o}]\rho_{\vec{n}-\vec{e}_k}(t).\nonumber\\
\end{eqnarray}



\begin{thebibliography}{99}
\bibitem{einstein1} A. Einstein, B. Podolsky, and N. Rosen, Phys. Rev. A \textbf{47}, 777 (1935).
\bibitem{bell1} J. S. Bell \emph{Speakable and Unspeakable in Quantum Mechanics} (Cambridge University Press, Cambridge, 1987).
\bibitem{nielsen1} M. A. Nielsen and I. L. Chuang, \emph{Quantum Computational and Quantum Information} (Cambridge University Press, Cambridge, 2000).
\bibitem{taylor1} P. L. Taylor and O. Heinonen, \emph{A Quantum Approach to Condensed Matter Physics} (Cambridge University Press, Cambridge, 2002).
\bibitem{modi1} K. Modi, A. Brodutch, H. Cable, T. Paterek, and V. Vedral, Rev. Mod. Phys. \textbf{84}, 1655 (2012).
\bibitem{schrodinger1} E. Schr\"{o}dinger, Naturwissenschafen \textbf{23}, 807 (1935).
\bibitem{vidal1} G. Vidal, Phys. Rev. Lett. \textbf{91}, 147902 (2003).
\bibitem{horodechi1} R. Horodechi, P. Horodechi, M. Horodechi, and K. Horodechi, Rev. Mod. Phys. \textbf{81}, 865 (2009).
\bibitem{amico1} L. Amico, R. Fazio, A. Osterloh, and V. Vedral, Rev. Mod. Phys. \textbf{80}, 517 (2008).

\bibitem{grover1} L. K. Grover, Phys. Rev. Lett. \textbf{79}, 325 (1997).
\bibitem{ahn1} J. Ahn, T. C. Weinacht, and P. H. Bucksbaum, Science \textbf{287}, 463 (2000).
\bibitem{knill1} E. Knill and R. Laflamme, Phys. Rev. Lett. \textbf{81}, 5672 (1998).
\bibitem{lanyon1} B. P. Lanyon, M. Barbieri, M. P. Almeida, and A. G. White, Phys. Rev. Lett. \textbf{101}, 200501 (2008).

\bibitem{ollivier1} H. Ollivier and W. H. Zurek, Phys. Rev. Lett. \textbf{88}, 017901 (2001).
\bibitem{zurek1} W. H. Zurek, Phys. Rev. A \textbf{67}, 012320 (2003).
\bibitem{henderson1} L. Henderson and V. Vedral, J. Phys. A \textbf{34}, 6899 (2001).

\bibitem{werlang} T. Werlang and G. Rigolin, Phys. Rev. A \textbf{81}, 044101 (2010);
T. Werlang, C. Trippe, G. A. P. Ribeiro, and G. Rigolin, Phys. Rev. Lett. \textbf{105}, 095702 (2010);
T. Werlang, G. A. P. Ribeiro, and G. Rigolin, Phys. Rev. A \textbf{83}, 062334 (2011).

\bibitem{dillenschneider1} R. Dillenschneider, Phys. Rev. B \textbf{78}, 224413 (2008);
M. S. Sarandy, Phys. Rev. A \textbf{80}, 022108 (2009);
L. Amico, D. Rossini, A. Hamma, and V. E. Korepin, Phys. Rev. Lett, \textbf{108}, 240503 (2012);
Y. Yao, H. W. Li, C. M. Zhang, Z. Q. Yin, W. Chen, G. C. Guo, and Z. F. Han, Phys. Rev. A \textbf{86}, 042102 (2012).


\bibitem{maziero1} J. Maziero, H. C. Guzman, L. C. C\'{e}leri, M. S. Sarandy, and R. M. Serra,
Phys. Rev. A \textbf{82}, 012106 (2010);
B. Tomasello, D. Rossini, A. Hamma, and L. Amico, Europhys. Lett. \textbf{96}, 27002 (2011);
B. Tomasello, D. Rossini, A. Hamma, and L. Amico, Int. J. Mod. Phys. B \textbf{26}, 1243002 (2012);
B. Q. Liu, B. Shao, J. G. Li, J. Zou, and L. A. Wu, Phys. Rev. A \textbf{83}, 052112 (2011);
Y. C. Li and H. Q. Lin, Phys. Rev. A \textbf{83}, 052323 (2011).

\bibitem{allegra1} M. Allegra, P. Giorda, and A. Montorsi, Phys. Rev. B \textbf{84}, 245133 (2011).
\bibitem{chen} Y. X. Chen and S. W. Li, Phys. Rev. A \textbf{81}, 032120 (2010).
\bibitem{wang} C. Wang, Y. Y. Zhang, and Q. H. Chen, Phys. Rev. A \textbf{85}, 052112 (2012)

\bibitem{gu1} M. Gu, H. M. Chrzanwski, S. M. Assad, T. Symul, K. Modi, T. C. Ralph, V. Vedral, and P. K. Lam, Nature Phys. \textbf{8}, 671 (2012).
\bibitem{walther1} B. Daki\'{c}, Y. O. Lipp, X. S. Ma, M. Ringbauer, S. Kropatschek, S. Barz, T. Paterek, V. Vedral, A. Zeilinger, \v{C}. Brukner,
and P. Walther, Nature Phys. \textbf{8}, 666 (2012).

\bibitem{breuer1} H. P. Breuer and F. Petruccione, \emph{The Theory of Open Quantum Systems} (Oxford University Press, New York, 2002).
\bibitem{weiss1} U. Weiss, \emph{Quantum Dissipative Systems} (World Scientific Publishing Company, 2008).


\bibitem{yu1} T. Yu and J. H. Eberly, Phys. Rev. Lett.  \textbf{93}, 14 (2004);
T. Yu and J. H. Eberly, Phys. Rev. Lett.  \textbf{97}, 140403 (2006);
T. Yu and J. H. Eberly, Quantum Inf. Comput. \textbf{7}, 459 (2007);
T. Yu and J. H. Eberly, Science \textbf{323}, 5914 (2009).

\bibitem{leggett1} A. J. Leggett, S. Chakravarty, A. T. Dorsey, M. P. A. Fisher, A. Garg, and W. Zwerger,
Rev. Mod. Phys. \textbf{59}, 1 (1987).

\bibitem{bellomo1} B. Bellomo, R. Lo Franco, and G. Compagno, Phys. Rev. Lett. \textbf{99}, 160502 (2007).
\bibitem{bellomo2} B. Bellomo, R. Lo Franco, and G. Compagno, Phys. Rev. A \textbf{77}, 032342 (2008).
\bibitem{maniscalco1} S. Maniscalco, F. Francica, R. L. Zaffino, N. Lo Gullo, and F. Plastina, Phys. Rev. Lett. \textbf{100}, 090503 (2008).
\bibitem{werlang1} T. Werlang, S. Souza, F. F. Fanchini, C. J. Villas Boas, Phys. Rev. A \textbf{80}, 024103 (2009).
\bibitem{wang1} B. Wang, Z. Y. Xu, Z. Q. Chen, and M. Feng, Phys. Rev. A \textbf{81}, 014101 (2010).
\bibitem{franco1} R. L. Franco, B. Bellomo, S. Maniscalco, and G. Compagno, Int. J. Mod. Phys. B \textbf{27}, 1345053 (2013).

\bibitem{cao1} X. F. Cao and H. Zheng, Phys. Rev. A \textbf{77}, 022320 (2008).
\bibitem{lee1} C. K. Lee, J. Moix, and J. S. Cao, J. Chem. Phys. \textbf{136}, 204120 (2012).



\bibitem{tanimura1} Y. Tanimura and R. Kubo, J. Phys. Soc. Jpn. \textbf{58}, 101 (1989);
Y. Tanimura, Phys. Rev. A \textbf{41}, 6676 (1990);
Y. Tanimura and P. G. wolynes, Phys. Rev. A \textbf{43}, 4131 (1991);
Y. Tanimura, J. Phys. Soc. Jpn. \textbf{75}, 082001 (2006).

\bibitem{ishizaki1} A. Ishizaki and Y. Tanimura, J. Phys. Soc. Jpn. \textbf{74}, 3131 (2005);
A. Ishizaki and Y. Tanimura, J. Chem. Phys. \textbf{125}, 084501 (2006);
A. Ishizaki and Y. Tanimura, J. Phys. Chem. A \textbf{111}, 9269 (2007).

\bibitem{xu1} R. X. Xu, P. Cui, X. Q. Li, Y. Mo, and Y. J. Yan, J. Chem. Phys. \textbf{122}, 041103 (2005);
Y. J. Yan and R. X. Xu, Annu. Rev. Phys. Chem. \textbf{56}, 187 (2005);
Q. Shi, L. P. Chen, G. J. Nan, R. X. Xu, and Y. J. Yan, J. Chem. Phys. \textbf{130}, 084105 (2009).

\bibitem{arend1} A. G. Dijkstra and Y. Tanimura, Phys. Rev. Lett. \textbf{104}, 250401 (2010).
\bibitem{ma1} J. Ma, Z. Sun, X. G. Wang, and F. Nori, Phys. Rev. A \textbf{85}, 062323
(2012).
\bibitem{kuhr1} S. Kuhr, S. Gleyzes, C. Guerlin, J. Bernu, U. B. Hoff, S. Del\'{e}glise, S. Osnaghi, M. Brune, J. M. Raimond, S. Haroche, E. Jacques,
 P. Bosland, and B. Visetin, Appl. Phys. Lett. \textbf{90}, 164101 (2007).

\bibitem{fanchini1} F. F. Fanchini, T. Werlang, C. A. Basil, L. G. E. Arruda, and A. O. Caldeira, Phys. Rev. A \textbf{81}, 052107 (2010).
\bibitem{wang0} C. Wang  and Q. H. Chen,  arXiv: 1202.5817.

\bibitem{niemczyk1} T. Niemczyk, F. Deppe, H. Huebl, E. P. Menzel, F. Hocke, M. J. Schwarz, J. J. Garcia-Ripoll, D. Zueco,
T. H\"{u}mmer, E. Solano, A. Marx, and R. Gross, Nature \textbf{6}, 772 (2010).

\bibitem{solano1} J. Casanova, G. Romero, I. Lizuain, J. J. Garc\'{i}a-Ripoll, and E. Solano,
Phys. Rev. Lett. \textbf{105}, 263603 (2010).
\bibitem{solano2} D. Ballester, G. Romero, J. J. Garc\'{i}a-Ripoll, F. Deppe, and E. Solano,
Phys. Rev. X \textbf{2}, 021007 (2012).

\bibitem{chen1} Q. H. Chen, Y. Yang, T. Liu, and K. L. Wang, Phys. Rev. A \textbf{82}, 052306 (2010).


\bibitem{Fink}  J. M. Fink, M. G\"{o}ppl, M. Baur, R. Bianchetti, P. J. Leek, A. Blais, and A. Wallraff, Nature \textbf{454}, 315 (2008).
\bibitem{Deppe}  F. Deppe, M. Mariantoni, E. P. Menzel, A. Marx, S. Saito, K. Kakuyanagi, H. Tanaka, T. Meno, K. Semba, H. Takayanagi, E. Solano, and R. Gross,
Nature Physics \textbf{4}, 686 (2008).


\bibitem{ferraro1} A. Ferraro, L. Aolita, D. Cavalcanti, F. M. Cucchietti, and A. Ac\'{i}n, Phys. Rev. A \textbf{81}, 052318 (2010).


\bibitem{werner1} R. F. Werner, Phys. Rev. A \textbf{40}, 4277 (1989).
\bibitem{popescu1} S. Popescu, Phys. Rev. Lett. \textbf{72}, 797 (1994).
\bibitem{hagley1} E. Hagley, X. Maitre, G. Nogues, C. Wunderlich, M. Brune, J. M. Raimond, and S. Haroche, Phys. Rev. Lett. \textbf{79}, 1 (1997).

\bibitem{garraway4} L. Mazzola, S. Maniscalco, J. Piilo, K.-A. Suominen, and B. M. Garraway, Phys. Rev. A \textbf{79}, 042302 (2009).
\bibitem{garraway1} B. M. Garraway, Phys. Rev. A \textbf{55}, 3 (1997).
\bibitem{garraway2} B. J. Dalton, S. M. Barnett, and B. M. Garraway, Phys. Rev. A \textbf{64}, 053813 (2001).
\bibitem{garraway3} B. J. Dalton and B. M. Garraway, Phys. Rev. A \textbf{68}, 033809 (2003).
\bibitem{mazzola1} L. Mazzola, S. Maniscalco, J. Piilo, K.-A. Suominen, J. Phys. B \textbf{43}, 085505 (2010).





\end{thebibliography}
\end{document}